# A Minimum Reconfiguration Probability Routing Algorithm For RWA in All-Optical Networks

Mohan Kumar S[1] and Jagadeesha S N[2]

[1]Research Center, Jawaharalal Nehru National College of Engineering,
Shimoga, Karnataka, India
[2]Department of Electronics and Communication Engineering, PES Institute of Technology and Management, shivamogga. Karnataka, India

*ABSTRACT*

*In this paper, we present a detailed study of Minimum Reconfiguration Probability Routing (MRPR) algorithm, and its performance evaluation in comparison with Adaptive unconstrained routing (AUR) and Least Loaded routing (LLR) algorithms. We have minimized the effects of failures on link and router failure in the network under changing load conditions, we assess the probability of service and number of light path failures due to link or route failure on Wavelength Interchange(WI) network. The computation complexity is reduced by using Kalman Filter(KF) techniques. The minimum reconfiguration probability routing (MRPR) algorithm selects most reliable routes and assign wavelengths to connections in a manner that utilizes the light path(LP) established efficiently considering all possible requests.*

## 1. INTRODUCTION

In this paper, we have considered Minimum Reconfiguration Probability Routing algorithm for Routing and Wavelength Assignment (RWA) for Wavlength Interchange(WI) network. It has scalable architecture and has mesh like structure consisting of links having one or more fibers at each input port to output port in the optical domain. The challenging problem in these networks are Routing and Wavelength Assignment and controlling problems. In these problems, provision of connections, called lightpaths in a scalable architecture usually span multiple links. Hence, light path might be assigned to different links along its route. This process is called Routing and Wavelength Assignment. In this process, a lightpath shares one or more fibers links and different wavelengths. To establish a lightpath, a route should be discovered between source and destination, and suitable wavelength need to be assigned to that route. Some of the commonly used performance criteria for Routing and Wavelength Assignment are throughput and blocking probability[10]. There are many RWA algorithms proposed for this purpose with optimal solutions[10]. The main assumption of these algorithms are that, traffic volume is static for a long time period and these networks are reconfigured only to reflect changes in the long term traffic demand[1]. Although static demand has been reasonable a assumption for voice data communication. In current trends and future, data intensive networks are rapidly changing. Therefore dynamic RWA algorithms which support request arrivals and lightpath terminations at stochastic/randaom times are needed. Hence, predefined set of routes are searched in a predefined order to accommodate the request. Then the smallest index randomly selects a wavelength available on the route. If not, request is blocked. Usually one or more minimum hop routes are used and fixed order search is carried out without taking into account, the congestion on the link. In this paper, we present the comparative study and performance evaluation of Minimum Reconfiguration Probability Routing (MRPR) algorithm, over Adaptive Unconstrained Routing (AUR) and Least Loaded Routing (LLR) methods.

An Optical switch has ability to minimize the effects of failures on network performance by using a suitable routing and wavelength assignment method without disturbing other performance criteria such as network management and blocking probability[4]. The main challenge in the Wavelength Interchange (WI) and Wavelength Routing Network (WRN) are the provision of connections called lightpaths between the users of network. One of the Routing Wavelength Assignment method known as Minimum Reconfiguration Probability Routing algorithm is implemented [3][15]. Its performance comparison study over AUR and LLR are presented. This algorithm makes use of the network state information at the time of routing to find the optimum path. By choosing as much reliable router/links as possible in the RWA process, it is possible to minimize the mean number of light paths broken due to failure. However, considering only reliability characteristics, lightpaths may have to be routed on longer routes and blocking performance may be reduced significantly. For this reason, the algorithm is based on the joint optimization of the probability of reconfiguration due to router/link failures and probability of blocking for the future requests. Therefore effect of potential router/link failures and blocking probability is measured and also minimized without disturbing performance. Hence, lightpath arrival/holding time and failure arrival statistics collected for each link and router, as well as the network state information collected and updated using kalman filter methods at the time of request arrival are used in routing decisions. That is, the behavior of the network is predicted by current state information and statistics of the past, to assign the most reliable path to the lightpath requests[9]. The routing algorithms proposed so far for optical networks uses present state of the network[3]. Using this information most of authors proposed wavelength pair routing and wavelenght assigment function. An adaptive RWA algorithm makes use of the network state information at the time of routing to find the optimum path. Least loaded routing(LLR) and Adaptive Unconstrained Routing (AUR) are the examples for adaptive RWA process[3].

In the following section, we present a MRPR for routing and wavelength assignment for the most reliable route-wavelength pair for a light path request. It uses the lightpath request arrival statistics between the routers, and failure arrival statistics for links and Kalman filter employed on router/linker to predict network state and the reconfiguration probability of route-wavelength pairs, and hence selects the most reliable route-wavelength pair. Finally a comparative study and performance are presented in the section 3.0. The results and concluding remarks and directions for future work are presented in the section 4.0 and section 5.0.

## 2. MRPR IN WI NETWORKS

In Wavelength Interchange networks, wavelength routers have a full set of wavelength converters at the output ports and they are able to change the wavelength over every light path passing through it. Hence, blocking of requests leading to wavelength conficts, can be measured and reduced significantly[15]. Wavelength router can also change the wavelength, it optimally reduce the ligth path routing problem after wavelength assignment problem is solved.

The minimum reconfiguration probability routing algorithm is statistically predictive optimal routing and wavelength assignment algorithm because; it aims to assign the most reliable route-wavelength pair for a lightpath request.  It uses the lightpath request arrival and failure statistics information from the routers, and predicts the reconfiguration probability of route-wavelength pairs, and selects the most reliable route-wavelength pair. The algorithm starts by assigning a cost value to each network element (router/link) and a multiple shortest path computed, then an optimal shortest path is used as the most reliable path.  To find the shortest route with minimum reconfiguration probability of light path request for reliable path for share per node (SpN) network, we use the Bellman Ford algorithm to find the shortest path. A kalman filter is employed at each router/link to retrieve the state of network and minimum reconfiguration probability of light path request for reliable path reconfiguring for the WI network. The following two sections determine these costs, and its route, a lightpath through the network.

### 2.1 COST DETERMINATION

Let 'x' be a random variable representing failure inter-arrival times, $f_{ij}(x)$ be the probability density function of failure inter-arrival times for link (i, j), y be a random variable representing lightpath holding times and $h_{sd}(y)$ be the probability density function of lightpath times for the router/link pair from source to destination (s-d). It is possible to evaluate the probability of failure of lightpath of the link (i, j), between routers source 's' and destination 'd' using these parameters.

Let $P_{sd}^{ij}$ be the probability of failure of lightpath in link (i, j) between router source and destination.  It is given by:

$$P_{sd}^{ij} = \int_{y=-\infty}^{\infty} (\int_{x=-\infty}^{y} h_{sd}(y) f_{ij}(x) dy dx ) \tag{1}$$

It is assumed that the failure arrival method is memory-less, hence it is the probability of failure of lightpath of the link (i, j) in the interval (t, t+dt).  It will not fail prior to the time t, hence it is independent of time t.

Rewriting the equation (1).

$$P_{sd}^{ij} = \int_{y=-\infty}^{\infty} \left[ \int_{x=-\infty}^{y} f_{ij}(x)\, dx \right] h_{sd}(y) dy = \int_{y=-\infty}^{\infty} g(y) h_{sd}(y) dy \tag{2}$$

*Tchebycheff Inequality*[21], is a measure of the concentration of a random variable near its mean $\mu$ and its variance $v = \sigma^2$, and can be used to find the upper bound for equation (2).

$$P_{sd}^{ij}[x - \mu] \geq \varepsilon\} = \int_{-\infty}^{x-\mu} f_{ij}(x)dx + \int_{x+\mu}^{\infty} f_{ij}(x)dx \leq \frac{v}{\varepsilon^2} \tag{3}$$

Where ε>0 and the $f_{ij}(x)$ is symmetric around its mean value.

Hence,

$$g(y) = \int_{x=-\infty}^{y} f_{ij}(x)dx \leq \frac{V_{sd}^{ij}}{2(\mu_{sd}^{ij}-y)^2} \tag{4}$$

Using equation (4) and equation (2), we get

$$P_{sd}^{ij} \leq \int_{y=-\infty}^{\infty} \frac{V_{sd}^{ij}}{2(\mu_{sd}^{ij}-y)^2} h_{sd}(y)dy \tag{5}$$

The estimation of upper bound is as follows

The mean of $g(y)$ is given by

$$E\{g(y)\} = \int_{-\infty}^{\infty} g(y)h(y)dy \cong g(\mu) + g'(\mu)\frac{V}{2} \tag{6}$$

By using equation (6) in equation (2), we get

$$P_{sd}^{ij} \leq \int_{y=-\infty}^{\infty} \frac{V_{sd}^{ij}}{2(\mu_{sd}^{ij}-y)^2} h_{sd}(y)dy \cong \frac{V_{sd}^{ij}}{2(\mu_{sd}^{ij}-\mu_h^{sd})^2} + \frac{3V_{sd}^{ij}V_{sd}^{sd}}{2(\mu_{sd}^{ij}-\mu_h^{sd})^4} \tag{7}$$

After simplification, we get:

$$P_{sd}^{ij} \leq \frac{V_{sd}^{ij}}{2(\mu_{sd}^{ij}-\mu_h^{sd})^2}\left[1 + \frac{3V_h^{sd}}{(\mu_{sd}^{ij}-\mu_h^{sd})^2}\right] \tag{8}$$

From the above, the link $(i, j)$ failure probability is given by,

$$P_{sd}^{ij} = \begin{cases} 1 & \text{if link}(i,j) \text{ is fully occupied or broken} \\ \min\left\{1, \frac{V_{sd}^{ij}}{2(\mu_{sd}^{ij}-\mu_h^{sd})^2}\left[1 + \frac{3V_h^{sd}}{(\mu_{sd}^{ij}-\mu_h^{sd})^2}\right]\right\} & \text{otherwise} \end{cases} \tag{9}$$

And similarly, the router $j^{th}$ failure probability is given by

$$Q_{sd}^{ij} = \begin{cases} 1 & \text{if node } j \text{ has failed,} \\ \min\left\{1, \frac{V_f^{ij}}{2(\mu_{sd}^{ij}-\mu_h^{sd})^2}\left[1 + \frac{3V_h^{sd}}{(\mu_{sd}^{ij}-\mu_h^{sd})^2}\right]\right\} & \text{otherwise} \end{cases} \tag{10}$$

From the above method, the cost of using link $(i,j)$ and router $j$ for a lightpath from source to destination can be written as:

$$C_{sd}^{ij} = -\ln(1-P_{sd}^{ij}), \quad C_{sd}^{j} = -\ln(1-Q_{sd}^{ij}), \tag{11}$$

## 2.2 Routing decision for MRPR, LLR and AUR techniques

In the previous section $P_{sd}^{ij}$ and $Q_{sd}^{ij}$ represent the estimated failure probabilities. Then the $(1 - P_{sd}^{ij})$ and $(1 - Q_{sd}^{ij})$ gives the probabilities of links *(i, j)* and route *j* are considered for routing a lightpath between source s to destination d. Here, probabilities are estimated for each network element independently. Hence, finding the shortest the path between router s and destination d in terms of cost can be written as:

$$C_{sd}^{ij} = -\ln[(1 - P_{sd}^{ij})(1 - Q_{sd}^{ij})] \tag{12}$$

$$C_{ij} = C_{sd}^{ij} + C_{sd}^{j} \tag{13}$$

Suppose if all link cost are equal, then the AUR and LLR routing approach will be adopted for routing decision. The kalman filter method is well suited for distributed operation since, it is possible to build an algorithm based on distributed routing algorithm. In this study, routers collect the failure statistics for present state, future prediction, updation of network state, their adjacent links and collects the arrival statistics from the router. Hence, router can initiate the routing algorithm with parameters' $\mu_h^{sd}$ and $v_h^{sd}$.

Suppose if wavelength converters are employed in routers, a wavelength can be assigned to a lightpath randomly or statistically at each link. This approach is called statistically predictive and optimal wavelength routing. It is also called Minimum Reconfiguration Probability Routing. The two possibilities for wavelength selection are first fit or random. In first fit wavelength assignment, the smallest index wavelength is chosen. In random wavelength assignment, one of the available wavelengths can be selected. First fit assignment methods are reported to be better than random methods and it is used in most of RWA algorithms [19].

A Markov Decision Process is used in WI networks to estimate the average number of future rejections on a router. The total cost is the estimation of the Mean number future rejections requests in the WI network. By accommodating the MRPR algorithm in the router, the statistics related lightpath arrival/holding time, failure arrival statistics for each link and routers, network state information at the time of request arrival are used in routing decision.

After solving the routing problem, wavelengths on the links along the route can be assigned randomly. In order to find the route with minimum reconfiguration probability for a lightpath request, a simple auxiliary graph G= *(N,E)*,where nodes *N* represent the routers and each directed edge*(i,j)*, *E* represents the link *(i,j)* from router *i* to router *j*, is constructed. Then, cost of each edge*(i,j)* is set to:

$$C_{ij} = -\ln(1 - F_{ij}) - \ln(1 - F_j), -\ln(1 - R_{ij}) \tag{14}$$

Where, $F_{ij}$ is the probability of reconfiguration due to failure of the link *(i,j)* and $F_j$ is the probability of reconfiguration due to failure on router *j* for the lightpath to be routed and $R_{ij}$ is the probability of reconfiguration due to repacking on the link *(i j)*. In equation (14), Edge cannot be used which means that the link *(i,j)* has no free wavelength channel at the time of routing[3][15], if the lightpaths assigned are uniformly distributed over the wavelength in link *(i,j)* (i.e arrival rates of lightpaths at different wavelenghts are the same).

Wavelength converters are exclusively expensive devices and hence it is treated as routers or switches. Therefore, some router architectures required Share per Node (SpN) and share per link method. Share per node architectures (network) are limited numbers of converters, among the lightpaths, in these networks, light path requests arrive at each routers randomly, and not blocked, they stay on the network for a random amount of time. It is assumed that, each links/router may fail at random times, so failures are re-routed on new path [14]. Hence, most of authors proposed share a limited channel number based WI routers. Although MRPR can be employed for the WI architectures. MRPR routing method may cause blocking of future requests, which need wavelength conversion at the router. Therefore, we consider the repacking of lightpaths, which are using converters in a router. For this purpose, we use converter usage statistics such as mean arrival time rate and mean holding time of light paths. Establishing a light path between the routers source 's' and destination 'd' in SpN network can be constructed as follows.

- For each wavelength w=1..W and router n, create the nodes $i_{nw}$ and $o_{nw}$ which represent the input and output ports of the routers n at wavelength w respectively.
- For each wavelength v, w=1..W and routers n create the edges ($i_{nv}$, $o_{nw}$) corresponding to wavelength conversion in the router n,
- For each wavelength w=1..W and each link (m, n) from router m to router *n*, create the directed edges ($o_{nw}$, $i_{nw}$) which represent the channels at wavelength *w* in the link *(m, n)*.
- Create the nodes $r_s$ and $r_d$ corresponding to source and destination terminals for the lightpath
- For each wavelength w=1..W create the edges ($r_s$, $i_{sw}$) and ($o_{dw}$, $r_d$)

After construction of graph for a share per node to route a lightpath from a source to destination are illustrated as shown in Figure (2). The graph is constructed for v, w=1, 2,..W and each router n and m, the cost of edges ($r_s$, $i_{sw}$) and ($o_{dw}$, $r_d$) as shown in Figure (2). for w=1..W are set to zero. The cost of each edge ($o_{nw}$, $i_{nw}$) are set to:

$$C_{mnw} = \begin{cases} \infty & \text{no free channel at wavelength w on (n,m)} \\ -\ln(1-F_{mn}) - \ln(1-R_{mn}) & \text{other wise} \end{cases} \quad (15)$$

Cost of each edge ($i_{nv}$, $o_{nw}$), for v ≠ w,

$$X_{nvw} = \begin{cases} \infty & \text{no free converters in n or n has failed,} \\ -\ln(1-F_n) - \ln(1-R_n) & \text{other wise} \end{cases} \quad (16)$$

And the cost of each edge ($i_{nw}$, $o_{nw}$)

$$X_{n\mu} = \begin{cases} \infty & \text{n has failed} \\ -\ln(1-F_n) & \text{other wise} \end{cases} \quad (17)$$

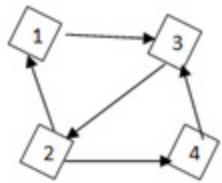

Figure 1 : A share per node network

Where $F_n$ and $F_{mn}$ are the probabilities of reconfiguration due to failure on the WI router n and link *(m, n)* respectively, $R_{mn}$ is the probability of reconfiguration due to repacking on the link *(m,n)* and $R_n$ is the probability of reconfiguration due to repacking for a lightpath using a converter on the router n which is obtained from the equation (15).

$$R_n = \frac{E(x,\rho)}{x.E(x_0,\rho)} \qquad (18)$$

Similarly, LLR and AUR methods are applied to WI network to find out the route and reliable path of network. Routing in optical networks is simply inherited from traditional circuit switching. Among all the routing algorithms, the Least Load Routing (LLR) is the most popular one in traditional circuit switched networks [22]. Naturally there were also attempts at studying LLR in the context of optical networks in; however, it has only been applied in two types of optical networks: a network with no wavelength conversion and a network with full wavelength conversion (i.e. every node has a wavelength converter). An LLR-based routing algorithm in optical networks with wavelength conversion neither has been formulated nor studied to the extent of our knowledge. In AUR algorithm, all paths between source and destination are considered in routing decisions. In LLR all paths between source and destination considered based on least cost and routing decision will be taken.

AUR is fundamentally different from LLR, in that, it is not limited to a set of predefined search sequence. The optimal lightpath is obtained by finding the shortest path in WI networks, both do not use convertor.

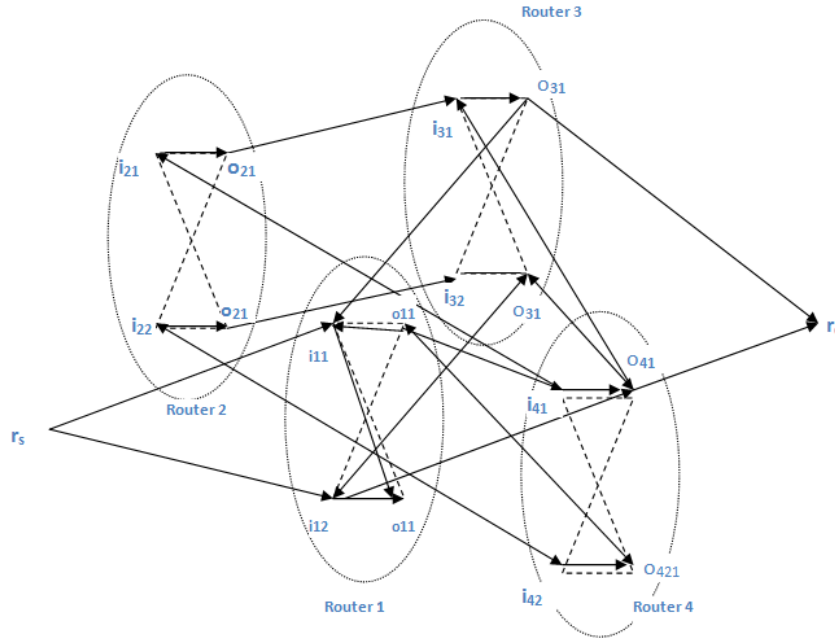

Figure 2 : Share per node graphical representation

## 3. PERFORMANCE EVALUATION

In order to evaluate the performance of network, we have considered the network shown in the Figure (3) having *6* routers and *13* links. All the links are unidirectional light paths and are established between the source to destination for each request. A Light path request arrives at a time of each request as a poisson process, and lightpat holding times are assumed to be exponential process. Each link is composed of single fiber length with three wavelength on each fiber link. We deploy kalman filter at link/routers to make use of the network state information at the time of routing to find the optimum path for the request. Light path requests arrive at the network as a poisson process of rate $\lambda_T$ and are uniformly distributed from source to destination over the routers and routing decisions are made randomly for a lightpath request. The lightpath holding times are exponentially distributed with unit mean. The blocked lighpath requests are assumed to be lost.

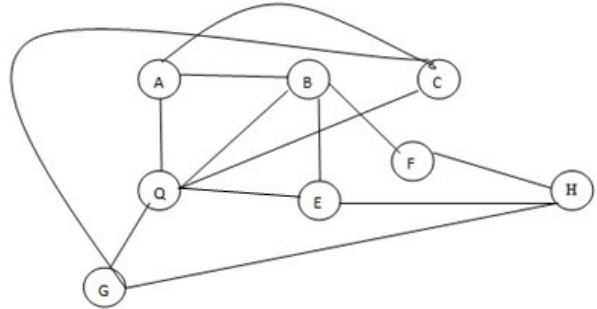

Figure 3: Assued topoogy having 6 routers and 13 links with single fiber

In order to assess the performance of MRPR, AUR and LLR, we consider two different types of routers characterised as reliable routers and un-reliable routers. The failures arrival at each reliable path or route, In all the computer simulations, the mean arrial of request rate and mean holding time are assumed to be poission processes of rates 0.0001 and 0.01 respectively. The service begin time, service elsaped time, serive ended and customer waiting time have been taken in to account in order to make routing decisions and reconfiguration of the network. We are interested in (1) blocking probability of a light path request (2) probability of re-routing due to link or router failure. We also measure the performance under changing load and reliablity conditions. The objective function is given by

$$\lambda = \frac{\lambda_T H}{WL} \qquad (19)$$

Where $\lambda_T$ is the total network load, and $H$ is the average number of fibers per link per path. W is the number of wavelength per fiber and $L$ is the number of fibers used in WI network.

The performance of network purely depends on reconfiguration and blocking performance with different reliablity and load values. Hence, for each load and reliabilty ratio values, we carried out *10* simulations.

In order to compute repacking probability, we need to know the initial arrival rate of lighpath and load offered to network for each resource. It is very difficult to determine actual offered load in the network, hence, we use light path arrival rates on each resource or service begin at network for link/router as its initial capacity (numbers of wavelength available on link/router). Then the number of occupied wavelengths are scanned every δ times and offered loads are updated using ϕ kalman functions.

For i= 1 to $\phi$

$$n_r(i) = number\ of\ lightpaths\ using\ resource\ r,$$

$$I_r = \begin{cases} 0 & n_r(i) < c_r \\ 1 & n_r(i) = c_r \end{cases} \tag{20}$$

$$N_r = \sum_{i=1}^{\phi} n_r(i)\Big/\phi, \quad B_r \sum_{i=1}^{\phi} I_r(i)/\phi \Rightarrow \quad \lambda_r = \frac{N_r}{h(1-B_r)} \tag{21}$$

Where $n_r$, $B_r$ and $\lambda_r$ are the number of lightpaths using resource $r$, blocking on resource $r$ and light path arrival rate on the resource called offered load on resource and $h$ is the average lightpath holding time.

## 4. RESULTS AND DISCUSSION

The performance of MRPR algorithm has been evaluated and compared with the performance of AUR and LLR algorithms. These alogrithms have been computer simulated using GNU Octave tool and Ubuntu Operating system 14.04 version. We try to find the shortest path from 'A' to 'H' as shown in Figure (3). In order to demostrate effect of the blocking probability and reconfiguration probability on the performace of MRPR, AUR and LLR, routing and wavelength assignment methods under changing load are presented. The reliabilty ratio is fixed at 0.05356, which means the network has un-reliable routers.

The reconfiguration probability is defined as the probability that a lightpath is re-routed due to a router or link failure. Since MRPR algorithm distinguishes between the available paths according to the failure statistics of the routers and links, and lightpath arrival statistics of routers, different failure arrival rates and variances are used for links and routers. For this purpose, some of the links/routers are assumed to be unreliable compared to others, and failure arrival rates, $\lambda_f$, for unreliable and reliable network elements are taken as 1/1500 and 1/1000 respectively in simulations.

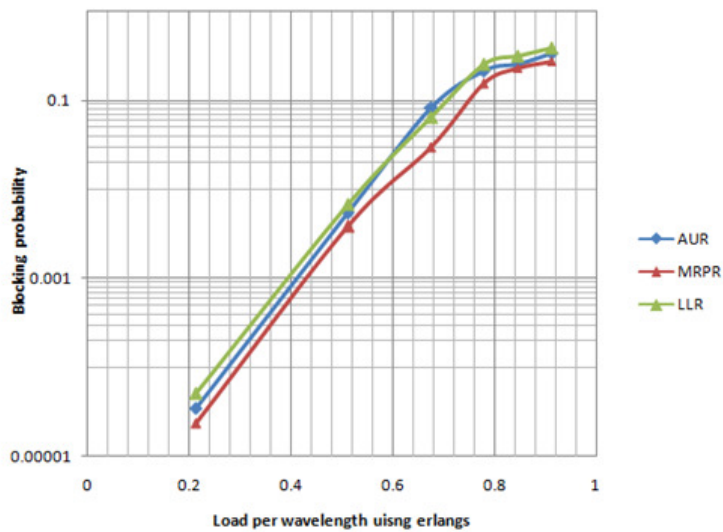

Figure 4: Performance in WI network whe reliablity ratio =0.05356 and wavelength per fiber =3 nos.

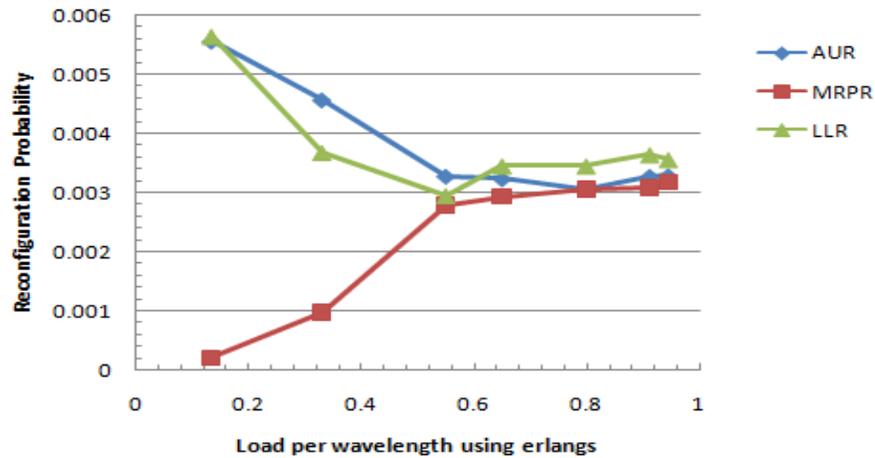

Figure 5: Performance in WI network when reliabilty ratio=0.05356 and wavelength per fiber=3nos.

In Figure (4) blocking probability as a function of load per wavelength obtained from the simulations are given. In these simulations, approximately 12% of routers are assumed to be unreliable. The results show that, from Figure (4) the blocking probabilities for AUR and MRPR algorithms are close to each other. This is an expected result, since AUR and MRPR are equivalent to each other, if all failure statistics are assumed to be the same. And also if the link costs in MRPR are taken as identical, then MRPR will be behave similar to AUR and LLR.

If reconfiguration probability is considered, MRPR algorithm performs 8% better than the AUR and LLR algorithm, and reconfiguration probabilities increase as the load per wavelength increases. This is due to the fact that, as the load per wavelength increases, utilization of fibers and routers increase and the number of lightpaths affected from a failure increases.

Table 1: Reconfiguration due to repacking

| No of channels | Mean failure inter-arrival time (μsec) | Mean inter-arrival time (μsec) | Mean holding time (μsec) | $R_p$ | $R_f$ | Cost |
|---|---|---|---|---|---|---|
| 1/3 | 1 | 3 | 2 | 0.00156 | **0.002781** | 1.100176 |
| 2/3 | 1 | 3 | 2 | 0.01853 | **0.002934** | 1.117317 |
| 3/3 | 1 | 3 | 2 | 0.33333 | **0.003045** | 1.504077 |
| 4/3 | 1 | 3 | 2 | 0.799 (above TH) | **0.003075** | INF |

In Figure (5) blocking probability and reconfiguration probability as a function of the ratio of unreliable routers and links are given. As expected, blocking performance of AUR and MRPR algorithms are close to each other. On the other hand, as the ratio of unreliable elements gets larger, the reconfiguration probability increases for both of the algorithms. Simulations show that MRPR algorithm performs approximately 13% better than AUR and 15% better than LLR algorithms. Table(1) shows the computed values for Bellman Ford algorithm to find the three shortest-paths that are encircled in Figure(2) using state space approach, Where $R_p$ is Probability of reconfiguration due to repacking, $R_f$ is Probability of reconfiguration due to failure at node and TH is Threshold level. As can be seen from the Table(1), for the channel three, the mean failure inter-arrival time is 1.0 μsec. Mean inter-arrival time 3.0 μsec, and mean holding time is 2.0μsec. The Probability of reconfiguration due to repacking is 0.33333, and Probability of reconfiguration due to failure is 0.003045 and the cost 1.504077. The proposed method to be any pratical value obviously, the process of taking statistics should not imply prohibitive costs. The cost is not prohibitive and Probability of reconfiguration is neglible.

In MRPR algorithm reliable paths between souce and destination are considered in the routing decisions. Since the statistics are used in conjunction with the present state information, it is naturally expected, that Kalman filter apporach, achieves optimaly better routing performance compared to earlier adaptive RWA algorithms.

## 5. CONCLUSION

In this paper, a detailed study of MRPR algorithm has been presented and its performance has been evaluated in comparision with AUR and LLR algorithms for the RWA problem in all optical networks. The study has been made considering the probability of reconfiguration due to failure of the link *(i, j)* and the probability of reconfiguration due to failure on router *j* for the lightpath to be routed and the probability of reconfiguration due to repacking on the link *(i j)*. The extensive computer simulations have reaveled that the MRPR algorithm is a better choice for solving the RWA problems. We suggest that further research in this direction is likely to find the over-head time taken to MRPR alogrithm for RWA problems.

## AUTHORS

**Mohan Kumar S.** received his Bachelor of Engineering in Computer Science & Engineering and Master of .Technology, in Networking and Internet Engineering, from Visvesvaraya Technological University, Belgaum, Karnataka, India respectively. He is currently pursuing a Doctoral Degree from Visvesvaraya Technological University, Belgaum, and Karnataka, India. At present he is working as Assistant Professor, Department of Information Science and Engineering M.S.Ramaiah Institute of Technology Bangalore. Karnataka. India (Affiliated to Visvesvaraya Technological University Belgaum). His research interest includes Communication and networking.

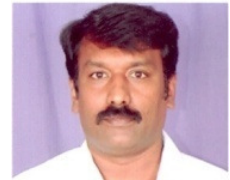

**Dr. Jagadeesha S N** received his Bachelor of Engineering., in Electronics and Communication Engineering, from University B. D. T College of Engineering., Davangere affiliated to Mysore University, Karnataka, India in 1979, M.E. from Indian Institute of Science (IISC), Bangalore, India specializing in Electrical Communication Engineering., in 1987 and Ph.D. in Electronics and Computer Engineering., from University of Roorkee, Roorkee, India in 1996. He is an IEEE member. His research interest includes Array Signal Processing, Wireless Sensor Networks and Mobile Communications. He has published and presented many papers on Adaptive Array Signal Processing and Direction-of-Arrival estimation. Currently he is professor in the Department of Electronics and Communication Engineering, PES Institute of Technology and Management. (Affiliated to Visvesvaraya Technological University), Shivamogga, Karnataka, India.

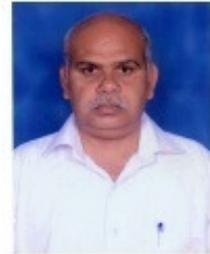